# Proposed high-power beta cells from MgAlB$_{14}$-type icosahedral-boron semiconductors

David Emin
Department of Physics and Astronomy
University of New Mexico
Albuquerque, New Mexico 87131, USA

Beta cells generate electric power as carrier-producing beta irradiation from incorporated radioisotopes bombard a series of *p-n*-junctions. However, radiation damage to the semiconductors commonly used in solar cells limits beta cells to extremely weak irradiations that generate concomitantly miniscule electric powers, e.g. micro-Watts. By contrast, beta cells that generate many orders-of-magnitude larger powers are possible with icosahedral boron-rich semiconductors since their bombardment-induced atomic displacements spontaneously self-heal. Furthermore, substitutions for Mg and Al atoms of icosahedral-boron-rich semiconductors based on the MgAlB$_{14}$ structure can produce *p-n* junctions as electron transfers from doping-induced interstitial extra-icosahedral atoms convert some normally *p*-type materials to *n*-type. Moreover, electron-phonon interactions of the resulting readily displaceable interstitial cations with charge carriers foster their forming large polarons. Oppositely charged polarons repel one another at short range. These repulsions suppress the recombination of *n*-type with *p*-type polarons thereby increasing the beta-cell efficiency. All told, use of these icosahedral boron-rich semiconductors could enable beta cells with electric powers that are many orders of magnitude larger than those of existing beta cells. This development opens a new avenue for generating electricity from nuclear decays.

## I. INTRODUCTION

The utility of terrestrial solar cells as primary power sources is challenged by the solar flux's capricious variability arising from clouds, humidity, dirt and sand and its predictable daily and seasonal variations. By contrast, energy fluxes from products of nuclear decays are steadier and predictable.

Solid-state conversion of the energies of beta-rays into electrical energy utilize beta cells, devices that are analogous to solar cells. Beta cells that employ conventional semiconductors are severely limited due to their rapid degradation due to radiation damage.[1,2] As a result, current beta-cells use only very weak beta-irradiation to generate extremely small electric powers (e.g. micro-Watts).

The combination of three distinctive properties of the small subset of icosahedral boron-rich semiconductors based on the MgAlB$_{14}$ structure enable them to overcome this severe limitation. First, electron-bombardment damage to icosahedral boron-rich semiconductors spontaneously self-heals.[3,4] Second, some substitutions for Mg and Al interstitials within the network of boron icosahedra generate the moderate mobility *n*-type carriers needed for useful *p-n* junctions while smaller electron donation to the electron-deficient boron network yields its *p*-type carriers.[5] Third, interactions of the resulting displaceable interstitial cations with carriers foster their forming large polarons.[6,7] The short-range repulsion between oppositely charged polarons impedes their recombination thereby enhancing the carrier-separation efficiency of the beta cell's *p-n* junctions.[8,9]

As illustrated in Sec. 6, the resulting beta cells could generate electric powers and possess energy capacities that are many orders of magnitude larger than those of existing beta cells. Such devices would greatly increase the utility of solid-state conversion of nuclear energy to electrical energy.

## II. STRUCTURE AND SELF-HEALING OF ICOSAHEDRAL BORON-RICH INSULATORS

Figure 1 illustrates the structure of the simple icosahedral boron-rich insulators, B$_{12}$P$_2$ and B$_{12}$As$_2$.[4,10] Distinctively, these solids contain boron atoms that reside at the twelve vertices of icosahedra. Each of these atoms forms a conventional two-center bond with an atom external to its icosahedron. By contrast, the thirteen internal-bonding molecular orbitals of an icosahedron distribute charge about the centers of its twenty faces. A boron icosahedron has an affinity for two electrons since the sum of the 12 electrons donated to its external covalent bonds plus the $26 = 2 \times 13$ electrons needed to fill its internal bonding orbitals exceeds the $36 = 3 \times 12$ second-shell electrons provided by twelve boron atoms.

Massive high-energy bombardments with electrons and ions readily knock atoms of icosahedral boron-rich insulators from their equilibrium positions.[3,4] However, as shown in Fig. 2, the amorphization and defect clustering that occurs for other semiconductors is not observed for icosahedral boron-rich semiconductors. Rather, such damage apparently self-heals as bombardment-induced vacancies and interstitials recombine.[3]

This behavior is fostered by an icosahedron retaining its internal bonding electrons when an atom is knocked from it. The atom then exits as a cation as the icosahedron garners an additional negative charge. Indeed, icosahedra which are "degraded" by the loss of an atom remain structurally stable as they each accumulate an extra negative charge.[11] Self-healing occurs as degraded icosahedra recombine with rapidly diffusing small interstitial boron cations.[3]

### III. CHARGE CARRIERS IN SEMICONDUCTORS ARE EITHER FREE, SMALL POLARONS OR LARGE POLARONS

The adiabatic principle governs the state of a semiconductor's charge carriers.[6,7,9,12] As shown in Fig. 3, there are two distinct adiabatic solutions for a charge carrier in an isotropic covalent material with its short-range (e.g. deformation-potential) electron-phonon interaction. Either an electronic carrier remains free or it collapses into a severely localized self-trapped state thereby forming a small polaron. Unlike free carriers, small polarons generally move incoherently by low-mobility phonon-assisted hopping.

A different situation prevails in semiconductors having significant densities of displaceable ions. As illustrated in Fig. 4, with only the long-range electron-phonon interaction arising from an electronic carrier's Coulomb interactions with displaceable ions, the adiabatic treatment for an isotropic material only permits large-polaron formation. However, adding a sufficiently strong short-range component to the long-range electron-phonon interaction drives a large polaron's collapse into a small polaron.

A large polaron's self-trapped electronic carrier is bound within the potential well produced by carrier-stabilized shifts of surrounding ions' equilibrium positions from their carrier-free values. Since a large-polaron's motion is contingent on movements of these ions, it moves slowly with a very large effective mass. A coherently moving large polaron's very large effective mass usually ensures its weak-scattering by ambient phonons.[7,13] A large polaron's long scattering time then compensates for its large effective mass to produce a moderate mobility, ~1-100 cm$^2$/V-s at 300 K.

The minimum mobility for coherently moving charge carriers occurs when their mean-free-path falls to their de Broglie wavelength.[7,14] This minimum mobility then depends on the charge carrier's effective mass $m^*$, its charge $e$, the thermal energy $kT$, and Planck's constant $h$: $\mu_{min} \equiv eh/m^*kT$.[7] For a free carrier with an effective mass as large as the free-electron mass $m_e$ $\mu_{min} = 300$ cm$^2$/V-s at 300 K. By contrast, the huge effective mass of a large polaron $m^* >> m_e$ gives $\mu_{min} \sim 1$ cm$^2$/V-s at 300 K.

Large polarons are identified by the distinctive frequency dependence of their electrical conductivity.[7,15] Large-polarons' long scattering time restricts their Drude response to frequencies below those of the associated phonons. Meanwhile, excitation of a large polaron's self-trapped electronic carrier from the potential well within which it is bound produces a broad absorption band at frequencies above those of the associated phonons. The pseudo gap between these two features opens as lowering the temperature increases the Drude conductivity's scattering time.

Measurements of the static and high-frequency dielectric constants of condensed matter indicate the presence of displaceable ions. In covalent semiconductors, the ratio of the static to high-frequency dielectric constants $\varepsilon_0/\varepsilon_\infty$, is only slightly greater than one. By contrast, the displaceable ions of simple ionic solids, e.g. alkali halide crystals, generates dielectric-constant ratios of $\varepsilon_0/\varepsilon_\infty \cong 2$. Moreover, in materials with significant densities of especially displaceable ions, e.g. cuprate superconductors and hybrid organic-inorganic halide perovskite solar cells, $\varepsilon_0/\varepsilon_\infty \gg 2$.[16-21]

## IV. RECOMBINATION OF OPPOSITELY CHARGED POLARONS

Solar-cell-type devices operate by the electric field produced near the junction between $n$-type and $p$-type semiconductors separating photo-carriers before they can recombine. Conventional solar cells utilize covalent materials (e.g. Si or GaAs) whose high-mobility free carriers separate rapidly enough to forestall their rapid recombination. Unconventional solar-cells employ ionic semiconductors (e.g. metal and organometal halide perovskites) whose large polarons have much lower mobilities than those of high-mobility covalent solids.[21-26] However, recombination of oppositely charged large polarons as well as oppositely charged small polarons are suppressed by their mutual short-range repulsions.[7-9]

Figure 5 schematically depicts the net energy of an isotropic medium's oppositely charged polarons $E(s)$ as a function of their separation $s$. The net energy approaches that of two independent oppositely charged polarons, $-(E_{p+} + E_{p-})$, when their mutual separation greatly exceeds the sum of their radii, $R_{p+}$ and $R_{p-}$. As $s$ falls toward $R_{p+} + R_{p-}$, constructive interference of the equilibrium positions of intervening ions enhances the long-range binding of the two oppositely charged polarons. However, as $s/(R_{p+} + R_{p-})$ falls below unity the net binding energy of the oppositely charged polarons approaches zero as the Coulomb fields that displace the equilibrium positions of surrounding ions progressively cancel one another. At $s = 0$ the oppositely charged carriers merge into an exciton of energy $E_{ex}$. With no net charge, the exciton lacks the Coulomb field to displace equilibrium positions of distant surrounding ions.

A domain for which $\partial E(s)/\partial s < 0$ indicates a repulsive interaction between oppositely charged carriers. The condition $(E_{p+} + E_{p-}) > E_{ex}$ is sufficient, but not necessary, to produce such a repulsive interaction. The energy of independent electron and hole polarons separated by the distance $s$ and attracted by their mutual Coulomb interaction is:

$$E(s) = -\left[\frac{e^2}{4R_{p-}}\left(\frac{1}{\varepsilon_\infty} - \frac{1}{\varepsilon_0}\right) + E_{b-}\right] - \left[\frac{e^2}{4R_{p+}}\left(\frac{1}{\varepsilon_\infty} - \frac{1}{\varepsilon_0}\right) + E_{b+}\right] - \frac{e^2}{\varepsilon_0 s}. \quad (1)$$

The electron and hole polaron binding energies are each the sum of a long-range component and a short range component. Each of the long-range components explicitly depends on its respective polaron radius, $R_{p-}$ or $R_{p+}$, and on the material's static and high-frequency dielectric constants, $\varepsilon_0$ and $\varepsilon_\infty$. The short-range components for the electron- and hole-polaron binding energies are denoted by $E_{b-}$ and $E_{b+}$, respectively. The energy of the exciton is

$$E_{ex} = -\left(\frac{e^2}{2\varepsilon_\infty R_{ex}} + E_{b,ex}\right), \quad (2)$$

where $R_{ex}$ denotes the exciton radius and $E_{b,ex}$ represents the correction to the exciton's binding energy, the exciton's self-trapping energy, generated by its electronic carriers' altering the equilibrium positions of the atoms they contact.

In the covalent limit, $\varepsilon_0 \cong \varepsilon_\infty$, with $s \gg R_{ex}$:

$$E(\infty) - E_{ex} = \frac{e^2}{2\varepsilon_\infty R_{ex}} + E_{b,ex} - E_{b-} - E_{b+}. \quad (3)$$

In covalent semiconductors, oppositely charged conventional non-polaronic carriers attract one another, $E(\infty) > E_{ex}$, since then $E_{b-} = E_{b+} = 0$. However, oppositely charged polarons will experience a short-range *repulsion* when $E_{b-} + E_{b+}$ is sufficiently large.[7,9]

In semiconductors with exceptionally displaceable ions, $\varepsilon_0 \gg \varepsilon_\infty$, the exciton's Coulomb term tends to be overwhelmed by the long-range contributions to the polarons' binding energies:

$$E(\infty) - E_{ex} = -\frac{e^2}{4\varepsilon_\infty R_{p-}} - \frac{e^2}{4\varepsilon_\infty R_{p+}} + \frac{e^2}{2\varepsilon_\infty R_{ex}} - E_{b-} - E_{b+} + E_{b,ex}, \quad (4)$$

where $R_{ex} > R_{p-}$ and $R_{p+}$, since the Wannier exciton's reduced mass is less than the electronic effective masses of its electron and its hole. Furthermore, since polarons are charged while an exciton is neutral, the short-range contributions to polarons' binding energies tend to greatly exceed those for an exciton: $E_{b-} + E_{b+} \gg E_{b,ex}$. Thus, oppositely charged polarons in materials with especially displaceable ions, $\varepsilon_0 \gg \varepsilon_\infty$, tend to *repel* one another: $E(\infty) < E_{ex}$. This short-range repulsion suppresses recombination of oppositely charged polarons.

All told, in materials with exceptionally displaceable ions, $\varepsilon_0 \gg \varepsilon_\infty$, oppositely charged polarons can repel one another and polarons with the same signed charge can attract one another. In particular, large polarons of the same charge tend to merge to form large bipolarons in materials for which the ratio of static to high-frequency dielectric constants is exceptionally large, $\varepsilon_0/\varepsilon_\infty \gg 2$.[7,27,28] Moreover, the novel superconductivity of perovskite-based cuprates has been attributed to the ground-state of large bipolarons that condense into a liquid under the influence of their mutual phonon-assisted attraction.[29-32]

## V. ICOSAHEDRAL BORON-RICH SEMICONDUCTORS

Semiconductors generated by doping the wide-gap covalent icosahedral boron-rich insulators $B_{12}P_2$ and $B_{12}As_2$ were suggested for use in radiation-hard beta cells.[33] These are devices that are powered by irradiation with energetic electrons emitted during radio-isotopes' beta decay. Since these icosahedral boron-rich insulators have fourteen atoms per unit cell, their transport bands comprise many narrow energy bands. As a result, their free-carrier effective masses are much larger and their free-carrier mobilities are very much smaller than those of Si and GaAs, the semiconductors utilized in conventional solar cells. Furthermore, while *p*-type materials are easily realized, substitutional doping has yet to produce an *n*-type material.[34] All told, these features dampen the prognosis for producing efficient beta cells by substitutional doping of these covalent icosahedral boron-rich semiconductors.

By contrast, both *n*-type and *p*-type materials have been produced from icosahedral boron-rich semiconductors based on the $MgAlB_{14}$ structure. As schematically depicted in Fig. 6, the orthorhombic unit cell of $MgAlB_{14}$ structure contains four twelve-boron-atom icosahedra divided equally between two different orientations plus four pairs of extra-icosahedral boron atoms.[35,36] Each of these eight boron atoms forms two-center bonds with three different adjacent icosahedra. This structure's two metal atoms *partially* occupy sites within two inequivalent large extra-icosahedral open spaces. These materials can be doped by making substitutions for these metal cations.[5,37]

The formula $(R^{+Y})_y(M^{+X})_xB_{14}$ describes the $MgAlB_{14}$ structure having cations of metal R with valence $+Y$ and partial occupancy $y$ and cations of metal M with valence $+X$ and partial occupancy $x$. These metals donate $xX + yY$ electrons to a formula unit's structure. Two electrons are required to fill its icosahedron's internal bonding orbitals and two electrons are needed to produce a bond between its two non-icosahedral boron atoms. Thus, in this idealized situation, four electrons must be donated to each formula unit to just fill all of its bonding orbitals, thereby producing an insulator. Reducing this donation will generate *p*-type materials and increasing this donation will produce *n*-type materials.

Electronic transport measurements were recently reported on members of the series $(Y^{+3})_{0.56}(Al^{+3})_xB_{14}$.[5] The room-temperature dc resistivity is largest for $x = 0.53$, somewhat smaller for $x = 0.41$ and five orders-of-magnitude smaller for $x = 0.63$. Furthermore, measurements of the Seebeck coefficient for $x = 0.41$ indicate *p*-type conduction while those for $x = 0.63$ indicate *n*-type conduction. Quantitative estimates suggest moderate mobility *n*-type carriers and very low mobility *p*-type carriers.

These icosahedral-boron semiconductors have cations distributed among partially occupied sites off the boron-network. These cations' Coulomb interactions with charge carriers generate long-range components of their electron-phonon interactions. One can then envision *p*-type small polarons moving on the boron network and *n*-type large polarons moving between the cations outside of the boron network.

Thus, *p-n* junctions can be formed from these icosahedral boron-rich semiconductors possessing different values of *x*. Cations that are sufficiently displaceable to produce $\varepsilon_0/\varepsilon_\infty \gg 1$ necessarily generate a barrier to recombination of oppositely charged polarons. The suppression of recombination of *n*-type large polarons with *p*-type small polarons will enhance the efficiency of solar-cell-type devices based on these *p-n* junctions.

**VI. BETA CELLS OF ICOSAHEDRAL BORON-RICH SEMICONDUCTORS**

The solar intensity on earth varies greatly and capriciously with the weather and predictably with the time of day, latitude and season. For example, the average monthly solar intensity in London has a minimum of less than 0.003 W/cm$^2$ in January and maximum of about 0.02 W/cm$^2$ in July with a yearly average intensity of about 0.01 W/cm$^2$.[38] Each absorbed photon potentially generates an electron-hole pair.

By contrast, the power emitted from a radioisotope falls steadily with time as it decays. For example, the power emitted from materials containing beta-emitting $^{90}$Sr (e.g. Sr metal or SrTiO$_3$) falls from an initial value of ~2 W/cm$^3$ as it decays over its 28 year half-life.[1] Because the average energy of each bombarding beta electron is about 1 MeV, it induces between $10^5$ and $10^6$ electron-hole pairs.

Whereas a solar cell is powered by sunlight that impinges on it, the beta cell schematically illustrated in Fig. 7 is powered by the beta decay of radioisotopes contained within it. As such, a beta cell can employ linked *p-n* junctions amidst a distribution of radioisotopes. Shielding encases the beta cell to capture radiation (e.g. Bremsstrahlung) generated within it in addition to unabsorbed radiation from its radioisotopes. The shielding also conducts heat from this self-contained power source.

For example, if $^{90}$Sr comprises 1% of a 1 cm$^3$ beta cell, its initial power output, 0.02 W, is comparable to that of a 1 cm$^2$ solar cell in London at mid-summer with similar *p-n* junction efficiency. The net energy capacity of the 1 cm$^3$ beta cell, 0.8 W-year [= 0.02 W × 28 year ÷ ln(2)], is about 350 times that of a very good D-cell battery, 20 W-hr.

A beta-cell's output and capacity increase with the volume of effectively employed radioisotope. Thus, the power and net capacity of a cubic meter of beta cells would be a million times larger than that from a 1 cm$^3$ beta cell.

**VII. DISCUSSION**

A boron icosahedron retains its internal bonding electrons when it is degraded by the loss of a boron atom. As a result, a bombardment-induced degraded boron icosahedron takes an electron from the departing boron atom. This bombardment-induced damage then self-heals as interstitial boron cations spontaneously recombine with degraded icosahedra. Thus, bombardment-induced damage to icosahedral boron-rich semiconductors self-heals. As a result, the lifetimes of beta cells based on icosahedral boron-rich semiconductors will not be limited by bombardment-induced atomic displacements.

Beta cells require junctions between *p*-type and *n*-type icosahedral boron-rich semiconductors. Unfortunately, since icosahedral boron networks are electron-deficient covalent networks, icosahedral boron-rich semiconductors usually manifest *p*-type transport. Nonetheless, icosahedral boron-rich semiconductors based on the MgAlB$_{14}$ structure can be doped *p*- and *n*-type by altering the valences and partial occupations of metal cations in large holes between icosahedra.

Displaceable cations generate ratios of a material's static to high-frequency dielectric constants that greatly exceed unity, their values in covalent solids. Charge carriers' Coulomb interactions with displaceable cations also produce the long-range electron-phonon interactions that foster large-polaron formation. Large polarons are identified by their moderate mobilities and by the existence of pseudo-gaps in their distinctive frequency-dependent absorption spectra.[15]

Unlike the situation for conventional charge carriers in covalent semiconductors, an energy barrier impedes the recombination of oppositely charged polarons. This suppression of recombination can significantly enhance the carrier-generation efficiency in polaron materials including biological matter. Thus, beta cells that utilize $MgAlB_{14}$-type icosahedral boron-rich semiconductors for their *p-n* junctions offer the possibility of more efficient, longer lived and higher capacity beta cells than have heretofore been possible.

Indeed, remarkably high-efficiency solar cells that utilize metal and organometallic halide materials with the perovskite crystal structure have been discovered.[39] Their high efficiencies have been attributed to squelched recombination of moderate-mobility large polarons whose formation is promoted by these materials' very large ratios of static to high-frequency dielectric constants.[8,22-25] The absorption spectra expected of large polarons have even been observed.[26]

Finally, just as short-range repulsions between oppositely charged polarons impede their recombination, short-range attractions between polarons of like charge leads to their pairing as bipolarons.[7,9] In particular, materials with exceptionally large ratios of their static to high-frequency dielectric constants foster large-bipolaron formation.[27,28] The large dielectric-constant ratios of cuprate superconductors result from displaceable ions that reside beyond the $CuO_2$ layers within which normal conduction occurs.[16-18] The superconductivity of suitably doped perovskite-based cuprates has been ascribed to the ground-state of the liquid of large bipolarons generated by their mutual phonon-mediated mutual attraction.[29-32]

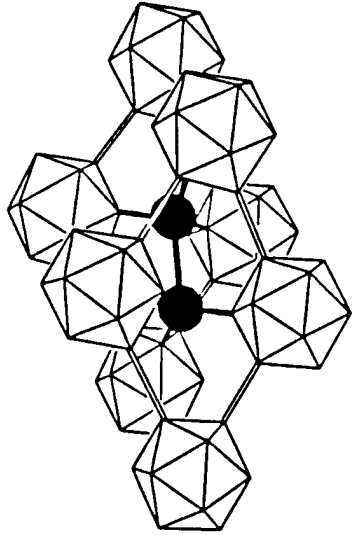

Fig. 1. The rhombohedral unit cell of the icosahedral boron-rich solids, $B_{12}P_2$ and $B_{12}As_2$, has boron-icosahedra at its corners and chains of two phosphorus or two arsenic atoms along its major diagonal.

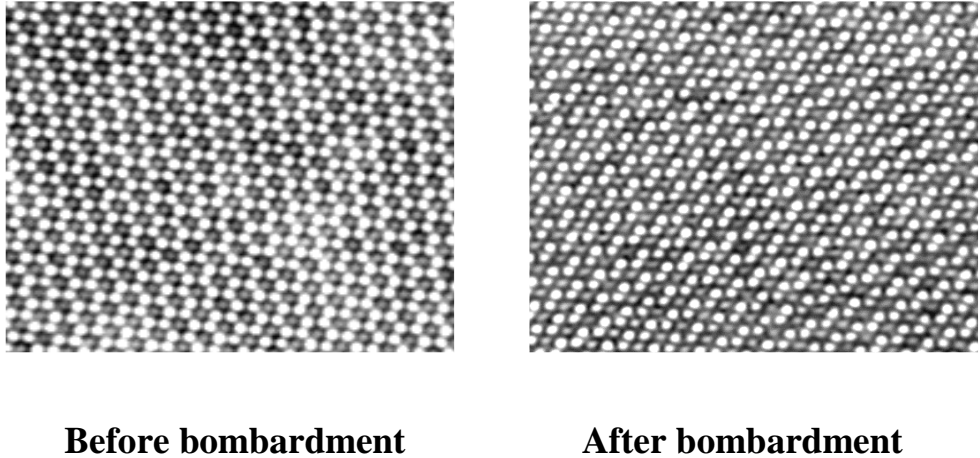

**Before bombardment**        **After bombardment**

Fig. 2. The HRTEM images (from L. Wang and R. Ewing in Ref. 4) of $B_{12}P_2$ before and after bombardment with a beam of 400 keV electrons whose intensity, $10^{18}$ electrons/cm$^2$-sec, is $10^6$ times that from undepleted $^{90}$Sr beta-emissions. No amorphization or defect clustering is seen after a net bombardment of $10^{23}$ electrons/cm$^2$, equivalent to that after 10,000 years of bombardment with constantly replenished $^{90}$Sr.

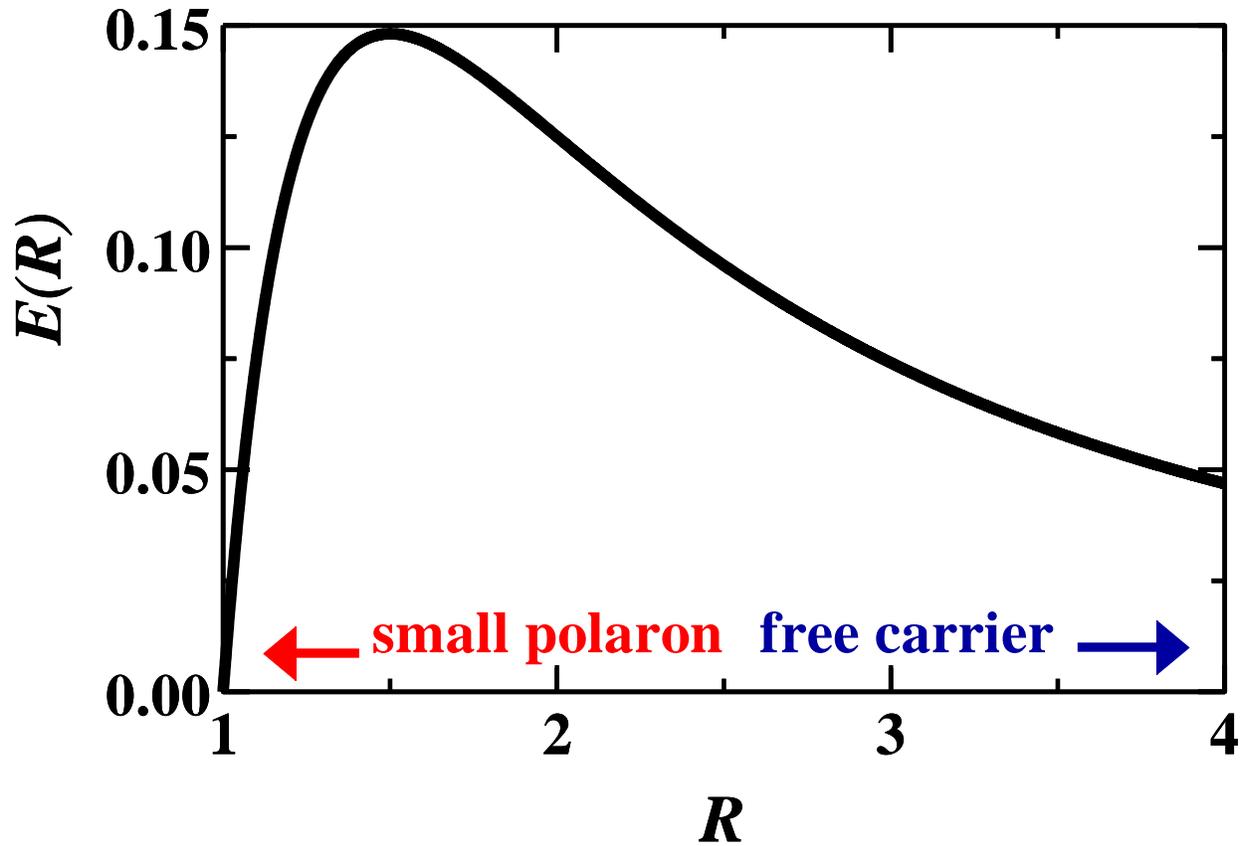

Fig. 3. The adiabatic energy $E(R)$ of a charge carrier with the short-range (deformation-potential-type) electron-phonon interaction of a covalent solid is plotted against the polaron radius $R$ in units of its minimum value (about an atomic radius). This function's two minima correspond to the system's two realizable states. The carrier either (1) collapses to a single site thereby forming a small-polaron or (2) expands to an infinite radius thereby remaining a free carrier.

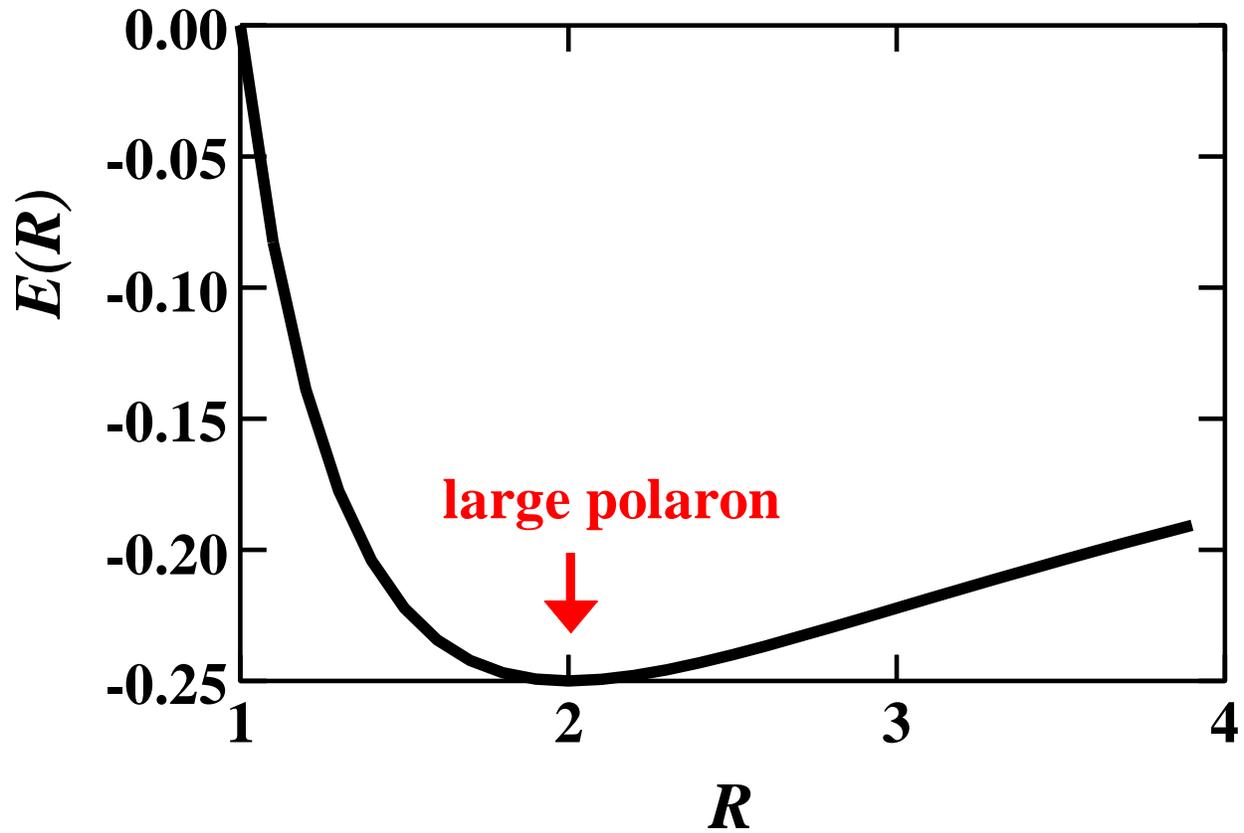

Fig. 4. The adiabatic energy $E(R)$ of a charge carrier with the long-range electron-phonon interaction that results from its Coulomb interactions with surrounding displaceable ions is plotted against the polaron radius $R$ in units of its minimum value (about an atomic radius). This function's solitary minimum corresponds to the carrier forming a large polaron.

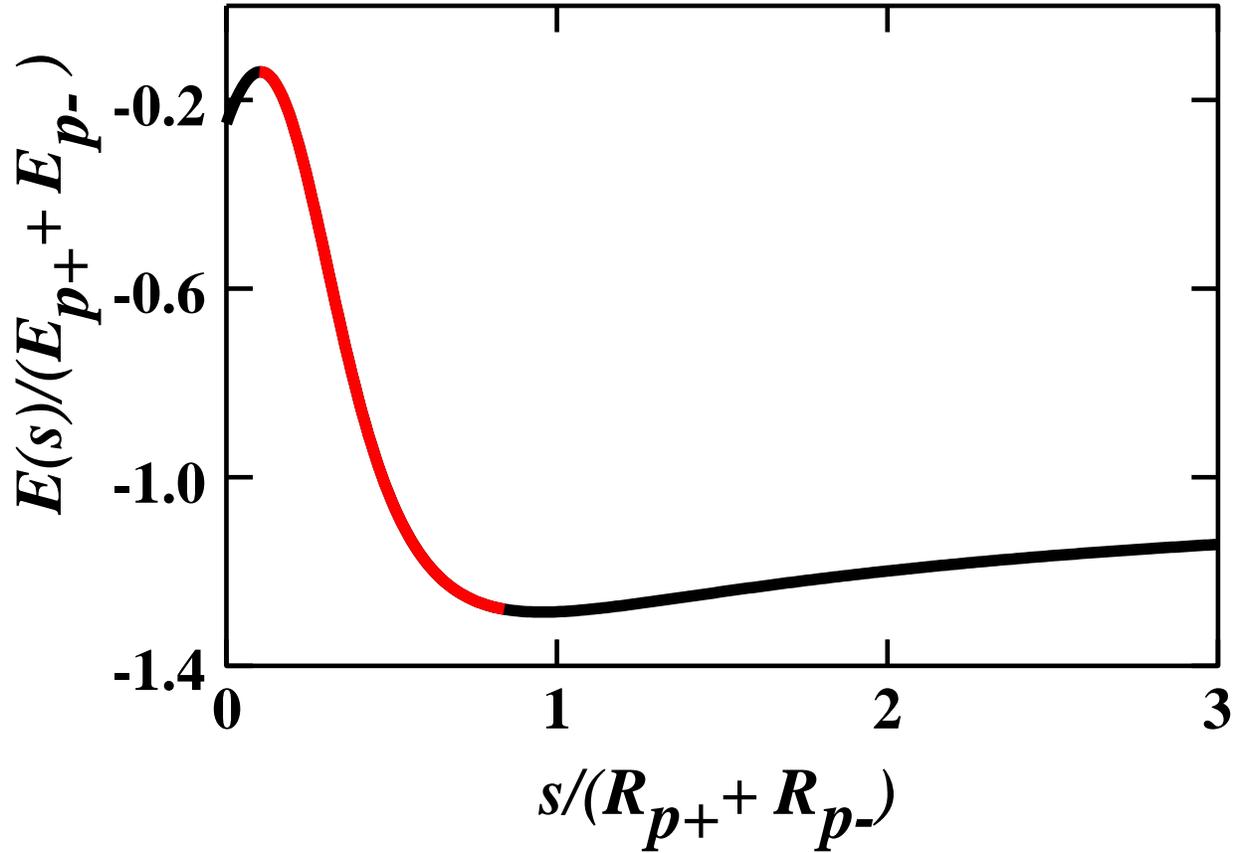

Fig. 5. The energy of two oppositely charged polarons $E(s)$ in units of $E_{p+} + E_{p-}$, the sum of their individual binding energies, is plotted against their separation $s$ in units of the sum of the two polaron radii, $R_{p+}$ and $R_{p-}$. As the inter-polaron separation is decreased from infinity, displacements of the equilibrium positions of intervening ions induced by the two polarons causes their net energy to decrease. However, the net inference of the ionic displacements changes from being constructive to being destructive as the two polarons begin to overlap. The surrounding ions then increasingly see the two polarons as being net neutral. The two polarons ultimately collapse into an exciton when $s = 0$. The red portion of $E(s)$ versus $s$ highlights the region in which the polarons exhibit a short-range mutual repulsion.

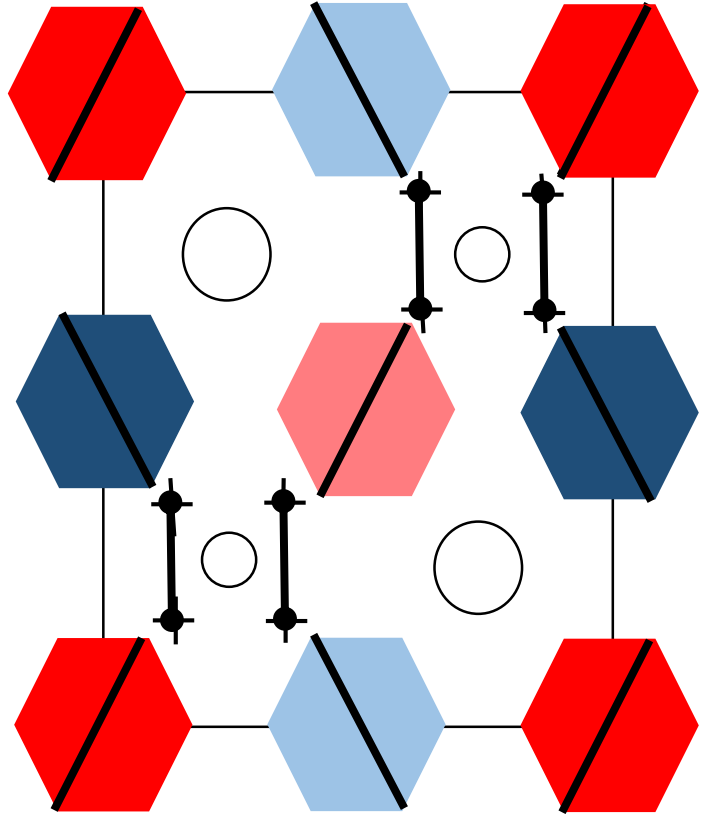

Fig. 6. The orthorhombic unit cell of MgAlB$_{14}$ encompasses four icosahedra divided between two different orientations. Modifying the pictorial representations of Refs. 35 and 37, hexagons' centers indicate the centroids of icosahedra while their colors and diagonal lines both indicate their orientations. Lightened colors indicate icosahedra that reside slightly below those of the cell's principal *yz* plane. Dots indicate the locations of pairs of extra-icosahedral boron atoms which are each bonded to three icosahedra. The large and small circles schematically indicate the idealized inequivalent partially occupied inter-icosahedral locations for Mg (2+) and Al(3+) cations.

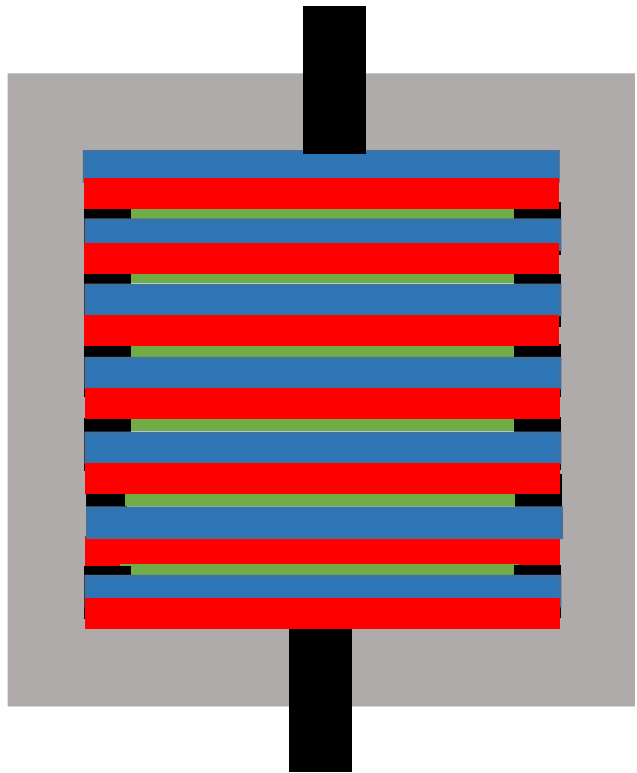

Fig. 7. Junctions between *p*-type material (blue) and *n*-type material (red) are connected in series by metal leads (black). This beta cell is powered by encapsulated volumes of radioisotope (green). The cell is encased in material (grey) that conducts heat to the outside while shielding it from radiation.